\documentclass[aps,singlecolumn,prd,epsf]{revtex4}
\usepackage{amssymb}
\usepackage{graphicx}
\newcommand{\be}{\begin{equation}}
\newcommand{\ee}{\end{equation}}
\newcommand{\bea}{\begin{eqnarray}}
\newcommand{\eea}{\end{eqnarray}}

\newcommand{\lpl}{\l_{\rm Pl}}

\begin{document}





\title{INFLATIONARY COSMOLOGY AND OSCILLATING UNIVERSES IN LOOP
QUANTUM COSMOLOGY}

\author{DAVID J. MULRYNE, N. J. NUNES, REZA TAVAKOL and JAMES E. LIDSEY}
\affiliation{Astronomy Unit,
School of Mathematical Sciences,
Queen Mary, University of London,
London E1 4NS, U.K. \\
d.j.mulryne@qmul.ac.uk
}






\begin{abstract}

\noindent We study oscillatory universes within
the context of Loop Quantum Cosmology.
We make a comparative study of 
flat and positively curved universes sourced by scalar fields
with either positive or negative potentials. We investigate 
how oscillating universes can set the initial conditions 
for successful slow-roll inflation, while
ensuring that the semi-classical bounds are satisfied.
We observe rich oscillatory dynamics with negative potentials,
although it is difficult to respect the semi-classical
bounds in models of this type. 
\end{abstract}


\maketitle

\section{Introduction}

The study of cyclic/oscillatory universes  
has a long history. One of the primary 
motivations for such a study was to circumvent the
need for initial conditions in cosmology. This was, however, shown
to be extremely difficult to achieve within
the context of general relativity (GR), without
encountering singularities.

Recent developments in String/M-theory 
inspired braneworld models
have revived hopes for the possibility
of constructing such universes 
(see e.g. Ref. \cite{kanekar-etal2001}).
One such attempt is the so called
cyclic/ekpyrotic scenario (see Ref. \cite{cyclic} and references therein).
Despite the attractiveness of this possibility,
a successful (non-perturbative) treatment of bounces
within the context of M-Theory is still lacking.

More recently such universes have been studied in the context
of loop quantum gravity, which at present is the main
background independent and non-perturbative
candidate for a quantum theory of gravity
(see for example Refs. \cite{rovelli98} and \cite{thiemann02}).
To date, cosmological applications
have focused on minisuperspace/midisuperspace settings
with a finite number of degrees of freedom. This program is
referred to as Loop Quantum Cosmology (LQC).
In this context the evolution of the universe
can be divided into 3 distinct phases:
an initial high energy and high curvature quantum
phase, described
by a (or a set of) difference equation(s);
an intermediate semi-classical phase,
with continuous evolution equations modified
due to non-perturbative
quantization effects; and
finally,  a classical phase,
where the usual continuous cosmological
equations are recovered.
It has recently been shown that such
semi-classical effects can result in bouncing
FRW universes which avoid
singularities \cite{st03}. This has in turn led to the possibility
of non-singular cyclic universes in LQC settings \cite{lmnt,Loop_cyclic}.

In this paper we make a comparative study of recent results
concerning both spatially flat \cite{blmst}, and
positively curved oscillating universes \cite{lmnt},
with both positive and negative potentials. We extend these results
in a number of areas, in particular, in the case of
models with negative potentials. This latter class of models 
is of direct relevance to the ekpyrotic/cyclic models considered
recently.

\section{Effective Field Equations in Loop Quantum Gravity}

The semi-classical phase in isotropic LQC refers to the regime
where  the scale factor
lies in the range $a_i < a < a_*$, where $a_i \equiv \sqrt{\gamma}\lpl$,
$a_*  \equiv \sqrt{\gamma j/3}  \lpl$,
$\gamma = \ln 2/\sqrt{3}\pi \approx 0.13$ and
$j$ is a quantization parameter which must
take half integer values. Below the scale $a_i$, the discrete nature of
space-time becomes important,
whereas the standard classical cosmology is recovered
above $a_*$. The parameter $j$ therefore sets the
effective quantum gravity scale.
The modified Friedmann equation is given by
\be \label{friedeq}
H^2 \equiv \left(\frac{\dot{a}}{a}\right)^2 =
\frac{8\pi\lpl^2}{3} \, \left[ \frac{1}{2} \frac{\dot{\phi}^2}{D}  +
V(\phi)\right] -\frac{k}{a^2}\, ~,
\ee
where $k$ takes the values $0$ or $1$ for a flat or positively curved
universe, respectively, and the quantum correction factor $D(q)$ is defined by
\begin{eqnarray}
D(q) = \left(\frac{8}{77}\right)^6 q^{3/2}
\left\{ 7\left[(q+1)^{11/4} - |q-1|^{11/4}\right]
\right. - \left. 11q\left[(q+1)^{7/4}- {\rm sgn} (q-1)|q-1|^{7/4}
\right]\right\}^6 \,
\end{eqnarray}
with $q \equiv (a/a_*)^2$. As the universe evolves through
the semi-classical phase, this function
varies as $D \propto a^{15}$ for $a \ll a_*$, has a global
maximum at $a \approx a_*$, and falls monotonically to $D = 1$
for $a > a_*$.

The scalar field equation has the form
\be
\label{fieldeq}
\ddot{\phi} = - 3 H \left( 1- \frac{1}{3} \frac{d \ln D}{d \ln a} \right)
\dot{\phi} - D V' ,
\ee
where a prime denotes differentiation with respect to the scalar field.
Differentiating Eq.~(\ref{friedeq}) and substituting from Eq.~(\ref{fieldeq})
then gives
\be
\dot{H} =  -\frac{4\pi \lpl^2 \dot{\phi}^2}{D}
\left( 1-\frac{1}{6} \frac{d \ln D}{d \ln a}  \right) - \frac{k}{a^2} .
\label{raycheq}
\ee

We can immediately see two interesting features.  First, the
correction
to the scalar field equation (\ref{fieldeq})  causes the
frictional/anti-frictional
$\dot{\phi}$ term to change sign when ${d \ln D}/{d\ln a}$
passes through 3. Secondly,
an expanding universe naturally undergoes a period of
super-inflationary
expansion  ($\dot{H}>0$) \cite{martin1,martin_kevin_1},
when ${d \ln D}/{d\ln a}>6$ (assuming the curvature term to be
either zero or negligible).

An interesting  consequence of these features is that both a
collapsing positively curved universe \cite{st03,lmnt}, and a flat
universe sourced by a scalar field with a negative
potential \cite{lmnt,Loop_cyclic}, can undergo a non-singular bounce.
We can most clearly see how the bounce arises by reformulating
Eqs.~(\ref{friedeq}) and (\ref{fieldeq}) in the
standard form of the Einstein field equations
sourced by a perfect fluid \cite{lmnt}:
\begin{eqnarray}
\label{standard1}
H^2 &=& \frac{8 \pi \lpl^2}{3} \rho_{\rm eff} - \frac{k}{a^2} \,, \\
\label{standard2}
\dot{\rho}_{\rm eff} &=& - 3H \left( \rho_{\rm eff} + p_{\rm eff}
\right) \,,
\end{eqnarray}
where
\begin{eqnarray}
\label{densityf}
\rho_{\rm eff} &\equiv& \frac{1}{2} \frac{\dot{\phi}^2}{D} + V \,, \\
\label{pressuref}
p_{\rm eff} &\equiv& \frac{1}{2} \frac{\dot{\phi}^2}{D}
\left( 1- \frac{1}{3}   \frac{d \ln D}{d \ln a}
\right) -V \,,
\end{eqnarray}
define the effective energy density and pressure of the fluid,
respectively.
The effective equation of state, $w \equiv p_{\rm eff}
/\rho_{\rm eff}$, is given by
\begin{equation}
\label{wf}
w =-1 + \frac{2 \dot{\phi}^2}{\dot{\phi}^2 +2DV}
\left( 1- \frac{1}{6} \frac{d \ln D}{d \ln a} \right)\,.
\end{equation}

Hence, the LQC corrections to the cosmic dynamics
can be entirely parametrized in terms of the equation of state.
When $d \ln D /d \ln a >6$, the fluid represents `phantom' matter $(w < -1)$
that violates the null energy condition
$(\rho_{\rm eff} + p_{\rm eff} \ge 0)$, independent of the form of the
potential. During a collapse the energy density starts to decrease
when this condition is met, and in a positively curved universe, the
density term in the Friedmann equation is eventually balanced
by the growing curvature term, thereby
leading to a bounce. In the case of a flat universe sourced
by a scalar field with a negative
potential, the decreasing energy density instantaneously
vanishes as the positive kinetic energy component is balanced by
the negative potential.
In both cases, $H$ tending to zero necessarily leads to a bounce
because $\dot{H}>0$.

In the following sections we shall study the question of
initial conditions for inflation in both flat and positively curved universes,
with both positive and negative potentials and we will investigate
whether other interesting scenarios involving negative
potentials can be developed.

\section{Scalar Fields with Positive Potentials}

In this section we consider a semi-classical universe sourced by a scalar field
self-interacting through a positive potential.  We are particularly
interested in how the dynamics of the semi-classical region can set
the initial conditions for slow-roll inflation if the field is situated
at or near the minimum of its potential.
Our discussion applies for
an arbitrary potential apart from the weak assumptions
that it has a global minimum
at $V_{\rm min} (0)=0$ and is a positive-definite and monotonically
varying function when $\phi \ne 0$ such that $V''>0$. We shall also assume
that the field is initially located at the minimum of its potential.
For definiteness, we focus in this section on a massive field
with a quadratic potential, $V(\phi ) = m^2\phi^2/2$.

\subsection{Flat case}

For a flat universe the
anti-friction mechanism simply accelerates the scalar field up the
potential \cite{tsm03,blmst}. A major problem with this
mechanism is that during the anti-friction era the energy density of
the field increases. On the other hand,
the Hubble length must remain larger than
the limiting value of the scale factor consistent with the semi-classical
regime, i.e., $|H|a_i  < 1$. (This is roughly equivalent to the requirement
that energy scales remain below the Planck scale
and is referred to as the Hubble bound). Since this length
decreases during the period of anti-friction, we need to determine how far
the field can be moved before this condition is violated.  To achieve
$60$ e-folds of inflation for a quadratic potential,
$V= m^2\phi^2/2$, the field must be displaced by at least
$3\lpl^{-1}$ from the minimum of its potential at the onset of
inflation. It was recently
found \cite{blmst}, that
given an initial value $\dot{\phi}_{\rm init}$, the field could only
be moved as far as $2.4\lpl^{-1}$ or less (for sufficiently large $j$)
without violating the Hubble bound.

The behavior of the field can be understood by approximating the
evolution into two epochs. In the first, between $a_{\rm init}$ and
the end of the semi-classical regime $a_S$, the asymptotic form,
$D_{\rm approx}=(12/7)^{6}(a/a_*)^{15}$, is used. In the second with
$a > a_S$, we assume $D=1$. Here  $a_S$ is
defined as the value of the scale factor 
where $D_{\rm approx}$ first reaches unity.  The field is
assumed to be massless up to the turning point in 
its evolution, which is estimated to occur when 
$m^2 \phi_{t} ^2/2 \approx \dot{\phi}^2_{\rm t}$, where
the subscript $t$ stands for ``turning point''. Hence, neglecting the
potential in the Friedmann equation (\ref{friedeq}), we have $(d\phi/d\ln
a)^2 = 6D/8\pi\lpl^2$.
Integrating for $\phi$, and using the approximation above for $D$, we 
then obtain
%
%
\begin{equation}
\phi_t = \phi_{\rm init} + \left(\frac{6}{8 \pi \lpl^2} \right )^{1/2}
\left\{ \frac{2}{15}\left(\frac{12}{7}\right)^3 \left[
  \left(\frac{a_S}{a_*}\right)^{15/2}-
  \left(\frac{a_{\rm init}}{a_*}\right)^{15/2} \right] +
  \ln\left(\frac{a_t}{a_S}\right) \right\} \,.
\end{equation}
The explicit dependence on $a_t$ can be removed by using
the expression
\begin{equation}
\dot{\phi}_t = \dot{\phi}_{\rm init} \left( \frac{a_S}{a_{t}}
\right)^3 \left ( \frac{a_S}{a_{\rm init}} \right )^{12} \, ,
\end{equation}
which itself arises from using the first integral
\begin{equation}
\label{firstintegral}
\dot{\phi} = \dot{\phi}_{\rm init} \left(
\frac{a_{\rm init}}{a} \right)^{3}
\frac{D}{D_{\rm init}} \,,
\end{equation}
in both epochs, along with the form for 
$D_{\rm approx}$.  Finally, applying the
condition for the turning point, we arrive at the estimate
\begin{eqnarray}
\label{eq:phimax0}
\phi_t \exp \left [ \sqrt{12 \pi} \lpl (\phi_t-\phi_{\rm init}) \right ]
= \frac{\dot{\phi}_{\rm init}}{m} \frac{\sqrt{2}}{q_{\rm init}^6}
\left (\frac{7}{12} \right )^{24/5} 
           \exp \left [ \frac{6}{15}\left( 1- (12/7)^3
q_{\rm init}^{15/4} \right)
  \right ] \,.
\end{eqnarray}
We note that for a given initial 
value $q_{\rm init} = (a_{\rm  init}/a_*)^2$, the value of
$\phi$ when inflation
occurs is independent of $j$ and proportional to $\dot{\phi}_{\rm init}$.
This approximation was compared to numerical results \cite{blmst}, and
found to be in very good agreement.
We emphasize that this approximation only 
works for initial conditions such that
$a_{\rm init} \ll a_*$ because the approximation for $D$ involves 
that assumption.

\subsection{Positively curved case - massless scalar field}

Before considering the positively curved case with a positive potential, 
it is illuminating 
to consider the dynamics of such a universe sourced by a massless scalar field.
This was discussed in detail in Ref. \cite{lmnt}. We can, however, easily
explain the form that the dynamics will take by the following simple argument.
We know from the above discussion that a collapsing
positively curved universe will always bounce into a new expanding phase
no matter what form the potential takes, and
consequently, this bounce will still occur even for $V=0$.
We also know that an expanding, positively-curved, classical
cosmology sourced by a massless scalar field will rapidly 
undergo a re-collapse since the energy density scales as $\rho \propto a^{-6}$, 
and therefore redshifts
much more quickly than the curvature term in Eq.~(\ref{standard1}).
After the bounce, the universe expands and once it has entered
the classical phase a re-collapse will inevitably occur.
The process then repeats itself indefinitely.

Since the effective equation of state
(\ref{wf}) is independent of the field's kinetic energy
when $V=0$, we see that identical cycles must occur.
Furthermore,
Eq. (\ref{firstintegral}) implies that the field's kinetic energy
never vanishes during a given cycle since
the scale factor always remains finite. The value of the field
therefore increases or decreases monotonically with time.

\subsection{Positively curved case --- massive scalar field}

To understand how the cyclic dynamics is modified by the presence of
a potential, we recall that the equation of state can change from cycle
to cycle depending on the field's position on the potential.
Assuming that the potential is initially unimportant,
the dynamics is qualitatively similar to that described in section 3.2.
This implies that the kinetic energy of the field never
vanishes and a re-collapse always occurs.
However, since the field moves monotonically 
away from the minimum, the potential
becomes progressively more important with
each completed cycle and it must therefore become dynamically
significant after a finite number of
cycles have been completed. In general, a re-collapse 
occurs if the matter redshifts more rapidly than the curvature term.
This requires the strong energy condition to hold, $w>-1/3$, which in
turn implies $V<\dot{\phi}^2$.  As the
field moves up the
potential,  $w$ is pushed towards $-1$ and it becomes progressively
harder to satisfy the strong energy condition. 
Eventually, therefore, a cycle is reached where
$V>\dot{\phi}^2$ and the condition $w>-1/3$ is violated. This leads
to an epoch of slow-roll
inflationary expansion, with the field slowing down,
reaching a point of maximum displacement and rolling back down the potential.
For some initial conditions and choices of parameters,
this may occur during the first cycle, in which case
the scenario discussed for the flat model is recovered.

Thus far, we have discussed the cyclic dynamics for a given value of the
quantization parameter, $j$. In spatially flat models,
we saw that the maximum value attained by the field as it moves up the
potential is independent of $j$ given an initial ratio $a_{\rm
init}/a_*$ and $\dot{\phi}_{\rm init}$.
For positively-curved models, on the other hand, the
field moves further up the potential before the onset of
slow-roll inflation for {\em smaller} values of $j$.
This behavior can be quantitatively understood
by assuming that the total energy at the point of collapse is conserved
through the cycles and that the potential is dynamically insignificant
during the first cycle. The massless case, therefore, gives us a good
estimate of the total energy, with the evolution of this energy being given by
Eq.~(\ref{firstintegral}). Now, assuming the universe to be classical at
$\phi_{t}$, using the condition $V_{t} \approx
\dot{\phi}_{t}^2$ to estimate the point at which inflation starts, and the
relation $8 \pi \lpl^2 \rho_{t}/3 \approx 1/a_{t}^2$ to eliminate
$a_{t}$, one obtains
\begin{equation}
\label{eq:phimax1}
\phi_{t}^2 = \frac{1}{\dot{\phi}_{\rm init}}
\frac{2}{m^2} \left (\frac{8 \pi \lpl^2}{2} \right )^{-3/2}
\frac{D_{\rm init}}{q_{\rm init}^{3/2}}
\frac{1}{a_*^3} \,.
\end{equation}
This demonstrates that for the quadratic potential,
with a given initial $a_{\rm init}/a_*$, the value of $V$ when inflation
occurs is inversely proportional to $a_*^3$ and hence inversely
proportional to $j^{3/2}$. It also shows an inverse dependence on
$\dot{\phi}_{\rm init}$. It is worth pointing out that in deriving
Eq.~(\ref{eq:phimax1}), unlike in the flat case,
we have not used any approximation to the
general form of the correction factor $D(q)$, 
and consequently this expression is
still a good estimate of $\phi_t$ when $a_{\rm init} > a_*$.

As in the flat model, it is important to ensure that 
the Hubble bound is not violated. However, due to 
the inverse dependence of $\phi_t$ on $\dot{\phi}_{\rm init}$, 
this proves to be a much weaker constraint in the positively 
curved models. The field is moved up the potential 
in a series of small `kicks' of anti-friction,  and can 
easily be moved sufficiently to drive $60$ e-folds of slow-roll 
inflation in the case of a quadratic potential \cite{lmnt}.

\begin{figure}[!t]
\centerline{
\includegraphics[width=8.5cm]{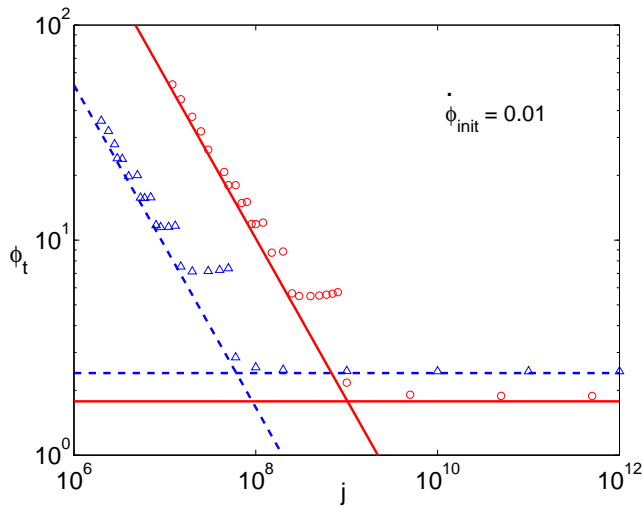}}
\caption[] {\label{figReza1.eps}Illustrating the dependence of the maximum
value of the scalar field, $\phi_t$,  on the quantization 
parameter $j$ for the quadratic potential
$V(\phi) = m^2 \phi^2/2$ with $m = 10^{-6}\lpl^{-1}$. 
The horizontal lines represent the estimated value of $\phi_t$
derived from Eq. (\ref{eq:phimax0}) and the tilted
lines represent the estimate (\ref{eq:phimax1}). The solid lines
correspond to the initial conditions $a_{\rm init}/a_* =0.7$ and 
the dashed lines to $a_{\rm init}/a_* = 0.5$. Circles and triangles 
represent the actual values obtained by numerically 
integrating the equations of motion for
$a_{\rm init}/a_* = 0.7$ and $a_{\rm init}/a_* = 0.5$.
}
\end{figure}

In Fig.~\ref{figReza1.eps}, we illustrate the dependence of the maximum value
attained by the field before turnaround on the value of the 
quantization parameter $j$ and the ratio $a_{\rm init}/a_*$
for a given initial value of the kinetic
energy $\dot{\phi}_{\rm init}$. 
The horizontal lines represent the
estimated value of $\phi_t$ extracted from Eq.~(\ref{eq:phimax0}) and the
lines with negative slope represent the estimate (\ref{eq:phimax1}).
Circles and triangles correspond to the actual values obtained by numerically
integrating the equations of motion. When many oscillations occur before
the onset of inflation, Eq.~(\ref{eq:phimax1}) yields a good estimate of
$\phi_t$. However, if the field is moved far enough in the first period of 
anti-friction for the strong energy condition to be violated, 
no oscillations occur and the situation is equivalent to the flat
model. In this case, Eq.~(\ref{eq:phimax0}) gives a good estimate of $\phi_t$.
The points where the lines cross in Fig.~\ref{figReza1.eps}
show the transitions between the two regimes.  

The numerical results in Fig.~\ref{figReza1.eps} show a stepping behavior, with
a range of $j$ leading to roughly the same value of $\phi_t$,
before there is a drop
to the next value. This becomes more pronounced at higher $j$. This occurs
because our approximation relies on two estimates: firstly, that
$V \approx \dot{\phi}^2$ at the onset of inflation, and secondly, that
$8 \pi \lpl^2 \rho_{t}/3 \approx 1/a_{t}^2$ when the field reaches its
maximal value. Clearly, the two approximations can not be satisfied
simultaneously, since the first implies a violation of the strong
energy condition, whereas the second represents a turning point in the
expansion. In effect, since we assume that the potential becomes dynamically
significant only during the last cycle \cite{lmnt}, we are
carrying over the second condition from the {\em previous} cycle. Since
$8\pi \lpl^2 \rho_{t}/3 > 1/a_{t}^2$ for the cycle in which inflation occurs,
the second condition underestimates the amount of energy present and
consequently underestimates the change in the value of the field before it
turns around. A range of values of $j$ will give rise to the same number of
cycles: for the lowest value of $j$ in a given range the condition
$8\pi \lpl^2 \rho_{t}/3 \approx 1/a_{t}^2$ is a good approximation,
since a re-collapse is only just avoided, but for
subsequent values the underestimate becomes progressively
more severe. This gives rise to the steps which become more evident towards larger $j$.
This can be understood by recalling from Eq.~(\ref{eq:phimax1})
that $\phi_t \propto j^{-3/4}$, which implies $\Delta j
\propto j^{7/4} \Delta \phi$, thus showing that the steps in
$j$ are wider for larger $j$.

\section{Scalar Fields with Negative Potentials}

In the spatially flat setting, we have seen that a bounce can occur
when the universe is sourced by a scalar field self-interacting
through a negative potential.  It is important to investigate whether
interesting cosmological scenarios can be developed using negative
potentials.  For example, a negative exponential potential  
plays a key role in the cyclic/ekpyrotic scenario \cite{cyclic}.
Alternatively, bouncing cosmologies in which the maximal size 
of the universe increases with
each successive bounce could offer a possible resolution of the flatness and
horizon problems of standard, big bang cosmology
(see Ref. \cite{solvebounce} and references therein).

Here we look at the general dynamics of a cosmic bounce with a negative
potential. During the collapsing phase,  
once the scale factor falls sufficiently for the condition
$d \ln D /d \ln a >3$ to be met, 
the kinetic energy of the scalar field decreases due to the friction in
Eq.~(\ref{fieldeq}). This is eventually balanced by the negative
potential energy  and leads to a bounce.  Since the balance is achieved
before the kinetic energy reaches zero, the field continues to move in
the same direction during the bounce. This is similar to the dynamics 
of a positively curved model with a positive potential.  
After the bounce, a super-inflationary era ensues, and the 
standard scenario is recovered once the
scale factor exceeds $a_*$. The field subsequently decelerates, and
in the case where the potential is negative-definite, 
a re-collapse is initiated when the kinetic energy is 
again balanced by the potential. 
The qualitative behavior discussed above is then repeated indefinitely with
the field always moving in the same direction.
The behavior, however, is not symmetric
since it responds to the region of the potential where the field is
positioned.  Consider, for example, the case of 
an exponential potential, $V \propto -\exp(-\alpha\phi)$: if the
field is moving up the
potential (towards more positive values), then the maximum size attained by 
the universe increases with each cycle whereas the minimal size decreases. 
This occurs because the universe has progressively more time to evolve 
before the kinetic energy is canceled by the decreasing negative 
potential term, and as the potential asymptotes to zero, 
the maximal/minimal size of the universe stabilizes. Conversely, if the
field is moving towards more negative values, the reverse is true.

One question that naturally arises is 
whether this mechanism for realizing a non-singular 
bounce may be employed within the context of the 
cyclic/ekpyrotic scenario?   
This question has been asked
previously \cite{Loop_cyclic}, and the discussion given here is in agreement with the previous study.
There are three major issues that need to
be addressed.  Firstly, in the cyclic/ekpyrotic scenario,
the field is supposed to change direction at the bounce, but as we have
discussed,
this is difficult to achieve at the semi-classical level with the mechanism under discussion, although
this problem might be alleviated by introducing
a steep positive section to the
potential. (This would cause the field to slow down and then reverse its
direction of motion).

The second problem is more severe and is directly
related to Hubble bound $Ha_i< 1$ discussed above.
During a classical collapse,
the kinetic energy of the field increases due to the anti-friction in
the field equation (\ref{fieldeq}). The kinetic energy can start to fall only
when the condition $d \ln D /d \ln a >3$ is attained.
Consistency at the semi-classical level requires that
the Hubble bound must not be violated before this point
is reached. (This problem of Planck scale energies
has been raised previously \cite{Linde}.)
In the case of a negative exponential potential, there exists
a scaling solution in which the kinetic and potential energies
of the field vary
in proportion to one another. In principle, therefore,
the total energy density may remain below the Planck scale.
For the cyclic/ekpyrotic scenario, however,
the potential has an exponential form only over a
finite region or parameter space, and once the field evolves through this
region, the cosmic dynamics moves away from the
scaling solution. This implies that the kinetic energy is no longer
balanced by the negative
potential, and the total energy rapidly exceeds that of the Planck scale.

For a collapsing universe sourced by a massless
scalar field, it is possible to estimate when the bounce must
occur in order for the Hubble bound to remain satisfied throughout
the evolution. In particular, we can derive a lower  limit on the value of the
ambiguity parameter, $j$. If the collapse starts well into
the classical regime, it follows that $H_{\rm init} =0$, $D = 1$, and that
the initial kinetic energy of the field is
$\dot{\phi}_{\rm init}^2= 6/8\pi\lpl^2a_{\rm init}^2$.
The subsequent evolution of the field is determined from the relation
$\dot{\phi}^2 = \dot{\phi}_{\rm init}^2 (a_{\rm
init}/a)^6$ and, after substituting this expression into the Friedmann
equation (\ref{friedeq}) and imposing the
consistency condition $Ha_i< 1$, we require that
\be
\label{jconstraint}
j > \frac{3}{\gamma} \left(\frac{a_{\rm init}}{a_*}\right)^4  \, ,
\ee
where
$a_*$ has been employed as the point at which the energy density becomes
maximized. Clearly, if the size of
the universe is large compared with $a_*$ at the start of the collapse,
$j$ must be very large indeed if a bounce is to occur within the
validity of the semi-classical regime.
Although a realistic collapse will involve a massive (or more generally a
self-interacting) scalar field, the above estimate
highlights the difficulties that arise: even if the field equations
generically lead
to a bouncing cosmology, the bounce may occur in a regime where the
equations are no longer valid.

Finally, a third question that arises in developing a
consistent bounce is also related to the size of
the universe. For a consistent, semi-classical
treatment, it is important that the bounce does not
occur after the universe has collapsed beyond the minimal
scale consistent with the
semi-classical region, $a_i$, since
the semi-classical Friedmann equations are no longer valid below this scale.
In particular, this is relevant to any oscillating scenario
involving a negative exponential potential in a flat universe where the maximal
size of the universe increases with each successive cycle:
as the value of the field increases with each cycle, the
value of the scale factor at the bounce decreases and it is possible that
it may fall below the critical value $a_i$.

\begin{figure}
\centerline{
\includegraphics[width=8.5cm]{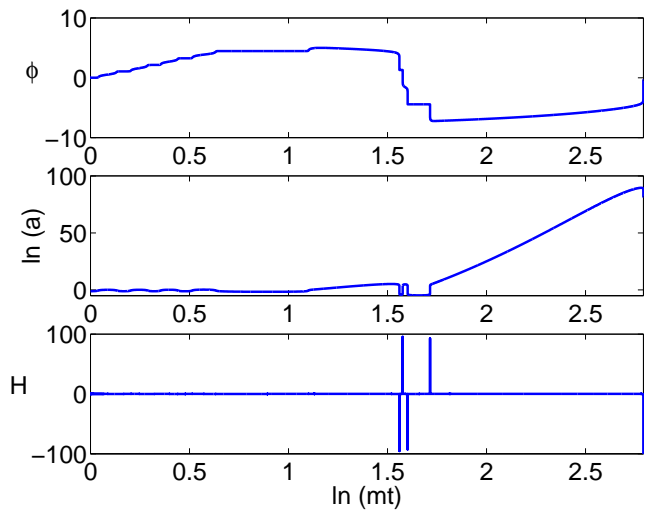}}
\caption[] {\label{figReza2.eps}
Time evolution of $\phi$, $\ln(a)$ and $H$ for a positively-curved universe
sourced by a quadratic potential $V(\phi)= C + m^2 \phi^2/2$, where
$m = 10^{-6} \lpl^{-1}$,
$j = 10$, $a_{\rm init}/a_* = 0.5$, $H_{\rm init} = 0$ and
$C = -10^{-11}\lpl^{-4}$. In the
figure,  the axes are labeled in Planck units.}
\end{figure}

Before concluding, it is also
worth considering the case where the
potential has a global minimum $V_{\rm min}$,
such that $V_{\rm min} < 0$, with only a region around the minimum that is
negative. One example of this class is a potential
of the form $V(\phi ) = C +m^2\phi^2/2$ for some
negative constant $C$.
In positively-curved universes, the mechanism described in section 3.2
enables the
field to work its way out of the negative region and
continue up the potential until the cycles are broken.
In Fig.~\ref{figReza2.eps} we illustrate the evolution of the scalar field,
the scale factor and Hubble ratio for a quadratic potential
$V=C+m^2\phi^2/2$ in a positively-curved universe. The figure shows the stepwise
evolution of the field up the
potential followed by a short period of inflation between $\ln (mt) =
[1,1.5]$. As the field rolls down the potential, it once again reaches the
negative section of the potential and this eventually
causes the universe to re-collapse. During the collapse the Hubble
parameter grows and is peaked around $a \approx
a_*$. Below $a_*$,
the potential prevents the scale factor from becoming singular and
the field resumes the same stepwise behavior, but now in the
opposite direction. This results in a further period of inflation.
For spatially
flat universes, the picture is similar
although the oscillations come to an end as soon as the
field reaches a positive region of the potential. Unfortunately, in the
flat case the resulting amount of inflation is
typically less that 60-efolds,
even for the quadratic potential. (More inflation is
possible but at the expense of violating the Hubble bound).
Furthermore, for such a potential in a flat universe,
there is the possibility that the potential becomes
positive during the collapsing phase, in
which case the universe can undergo no further bounces and
ultimately collapses into a big crunch.

\section{Concluding Remarks}

We have summarized and extended recent results
concerning the effects of loop quantum gravity corrections to the
cosmological evolution equations.
In particular, we have made a comparative study of
the ability of oscillatory universes
to set the initial conditions for successful slow-roll
inflation \cite{blmst,lmnt}.
We have considered both flat and positively
curved models sourced by a scalar field
with either a positive or negative self-interaction potential.
Negative potentials are of interest as they arise naturally
in string/M-theory motivated models such as the recently proposed
ekpyrotic/cyclic scenario \cite{cyclic}.
We find that the requirement for self-consistency
in the semi-classical regime leads to
strong constraints that any realistic model must satisfy.


\section*{Acknowledgments}

We thank Martin Bojowald and Param Singh
for their collaboration in some of the work reported here.
DJM and NJN are supported by PPARC.



\end{document}